\documentclass[letter]{jpsj2}

\title{%
Ordering of the pyrochlore Ising model with the long-range RKKY interaction
}

\author{%
Atsushige Ikeda and Hikaru Kawamura
}

\inst{%
Department of Earth and Space Science, Faculty of Science,
Osaka University, Toyonaka 560-0043,
Japan
}

\recdate{\today}

\abst{%
 The ordering of the Ising model on a pyrochlore lattice interacting via the long-range RKKY interaction, which models a metallic pyrochlore magnet such as Pr$_2$Ir$_2$O$_7$, is studied by Monte Carlo simulations. Depending on the parameter $k_F$ representing the Fermi wavevector, the model exhibits rich ordering behaviors. 
}

\kword{%
pyrochlore lattice, frustration, spin ice, RKKY interaction, Pr$_2$Ir$_2$O$_7$
}

\begin{document}
\maketitle

 Recently, the ordering of pyrochlore magnets consisting of corner-sharing network of tetrahedra has attracted much attention due to an intrinsic effect of geometrical frustration.  Intensive studies were performed in particular on ``spin-ice'' magnets, {\it e.g.\/}, Ho$_2$Ti$_2$O$_7$ and Dy$_2$Ti$_2$O$_7$, which possesses a strong Ising-like $<111>$ magnetic anisotropy and exhibits an intriguing glassy behavior analogous to the one associated with the proton ordering of water ice \cite{Harris,Ramirez,Bramwell}. In spin-ice magnets, the local spin configuration at each tetrahedron is known to be the so-called ``2-in 2-out'' configuration, which is stabilized by an effective ferromagnetic interaction. Indeed, this  ``2-in 2-out'' condition is reminiscent of the ``ice-rule'' condition familiar in the proton ordering of water ice.  While an earlier analysis on spin ice was made based on the nearest-neighbor ferromagnetic exchange model \cite{Harris}, the dominant interaction in real spin-ice magnets is the long-range dipolar interaction, which falls off as $1/r^3$ with distance $r$ \cite{Bramwell,review,Siddharthan,Hertog}. Thanks to extensive experimental and theoretical studies, the experimental results on spin ice magnets now seem to be relatively well understood on the basis of the dipolar spin-ice model, {\it i.e.\/}, the pyrochlore Ising model with the long-range dipolar interaction.

 Typical spin-ice magnets Ho$_2$Ti$_2$O$_7$ and Dy$_2$Ti$_2$O$_7$ are insulators. Ordering properties of the corresponding {\it metallic\/} pyrochlore magnets are then of special interest \cite{Bramwell}. Recently, Nakatsuji and collaborators reported that a metallic Ising-like pyrochlore magnet  Pr$_2$Ir$_2$O$_7$ exhibits an interesting behavior at low temperatures \cite{Nakatsuji}. The magnetism of this compound is borne by the localized moment of Pr$^{3+}$ which possesses a strong $<111>$ Ising-like magnetic anisotropy. Reflecting the metallic nature of the compound, the dominant magnetic interaction is expected to be the long-range RKKY interaction mediated by the Ir$^{4+}$ conduction electrons, rather than the dipolar interaction. Although the Curie-Weiss constant of Pr$_2$Ir$_2$O$_7$  determined from the high-temperature susceptibility measurement is antiferromagnetic $\theta _{\rm CW}\simeq -20$K, and the antiferromagnetic interaction usually prefers ``all-in all-out'' structure accompanying no frustration, the short-range magnetic order of this compound realized at lower temperatures appears to be ``2-in 2-out'' similarly to that of spin ice, characteristic of the effective ferromagnetic interaction.

 While  in spin-ice magnets the observed ferromagnetic Curie-Weiss temperature is entirely consistent with the ``2-in 2-out'' local spin structure \cite{Harris}, it remain a bit puzzling in Pr$_2$Ir$_2$O$_7$ how the ``2-in 2-out'' structure, which is usually an attribute of the ferromagnetic effective interaction, is realized under the antiferromagnetic $\Theta_{\rm CW}$. On decreasing the temperature, Pr$_2$Ir$_2$O$_7$ exhibits a spin-liquid-like behavior accompanied by the Kondo-like effect observed in its resistivity, until it eventually exhibits a weak spin freezing at a low temperature $T\simeq 0.12$K \cite{Nakatsuji}. 

 Under such circumstances, to better understand the ordering of metallic pyrochlore magnets, it would be desirable to study the nature of the magnetic ordering of the pyrochlore Ising model interacting via the long-range RKKY interaction, and compare it with the property of the dipolar spin-ice model. 

 We consider the Hamiltonian 
\begin{equation}
{\cal H} = -\sum _{ij} J(r_{ij}) \vec S_i\cdot \vec S_j,
\end{equation}
where $\vec S_i$ denotes a three-component unit vector along the $<111>$ direction located at the site $i$ of the pyrochlore lattice, pointing either parallel or antiparallel with the vector connecting  the site $i$ and the center of a tetrahedron.  The coupling between the spins at the sites $i$ and $j$ is the long-range RKKY interaction given by
\begin{equation}
J(r_{ij}) = -J_0a^3\left( \frac{\cos(2k_Fr_{ij})}{r_{ij}^3} - \frac{\sin(2k_Fr_{ij})}{2k_Fr_{ij}^4} \right), 
\end{equation}
where $r_{ij}$ is the distance between the sites $i$ and $j$, $k_F$ the Fermi wavevector, and the sum is taken over all spin pairs on the pyrochlore lattice. 

 The choice of the standard RKKY interaction (2) is justified when conduction electrons are sufficiently extended and are coupled with localized moments via the on-site exchange interaction. It is not clear at the present stage to what extent this assumption is satisfied in real metallic pyrochlore magnets such as Pr$_2$Ir$_2$O$_7$. If, {\it e.g.\/}, the Ir$^{4+}$ conduction electron has negligible amplitude at the Pr$^{3+}$ site, the d-f superexchange might play a role modifying the standard form of the RKKY interaction (2), although even in such a situation the qualitative feature of Eq.(2), {\it i.e.\/}, the long-range and the oscillating nature of the effective interaction, is still expected to hold. Furthermore, in understanding certain properties of real systems, the itinerant character of electrons, not captured by the spin model (1), might become important. In the following, with these possible limitations of the model in mind, we study the ordering properties of the RKKY Ising model (1)-(2), as a first step to understand the ordering properties of metallic pyrochlore magnets.   

 The model contains one parameter, the Fermi wavevector $k_F$. In case of Pr$_2$Ir$_2$O$_7$, experimentally determined carrier density  $n\simeq 4.13\times 10^{21}$ cm$^{-3}$ and the lattice constant $a\simeq 10.39\times 10^{-8}$ cm \cite{Nakatsuji} yield the dimensionless Fermi wavevector  $k_F\simeq 2\pi /1.218$ in units of $a^{-1}$.  Our length unit is taken to be the spacing of the cubic unit cell $a$. The lattice contains $N=16L^3$ spins (the cubic unit cell contains 16 sites). Periodic boundary conditions are applied.

%
%
%
%

 In order to study the ordering properties of the model, we perform Monte Carlo simulations on finite lattices, with $L$ in the range of $2\leq L\leq 6$.  In order to take account of the long-range nature of the RKKY interaction properly, we employ the Ewald sum technique, {\it i.e.\/}, the RKKY interaction is summed over to infinity under the condition that the images of the original lattice are repeated  periodically in all directions \cite{Hansen}. Typically,  each run consists of $5\times 10^4$ Monte Carlo sweeps, several independent runs being made to guarantee the statistical accuracy of the data.  Many of the data given below are taken at the ``experimental''  $k_F$-value,  $k_F=2\pi /1.218$, while other $k_F$-values are also investigated. 

 First, we concentrate on the ``experimental'' case of $k_F=2\pi /1.218$.  In Fig.1(a), we show the temperature dependence of the specific heat per spin. The data exhibit a sharp peak at around $T=T_c\simeq 16$ (in units of $J_0$). The peak grows rapidly with the size $L$, suggesting that the transition is of first-order. Indeed, as shown in the inset, the energy histogram, {\it i.e.\/}, the distribution function of the energy per spin $e$, exhibits a double-peak structure around $T_c$, which grows rapidly as $L$ increases.  In the inset of Fig.1(b), the inverse susceptibility $\chi ^{-1}=(\chi_{xx}+\chi_{yy}+\chi_{zz})^{-1}$ is shown as a function of temperature. The ordered state has no net magnetization. The estimated Curie-Weiss constant is negative, $\Theta_{\rm CW} \simeq -32$, apparently suggesting the occurrence of an {\it antiferromagnetic\/} effective interaction. 

\begin{figure}[ht]
\begin{center}
\includegraphics[scale=0.4,angle=-90]{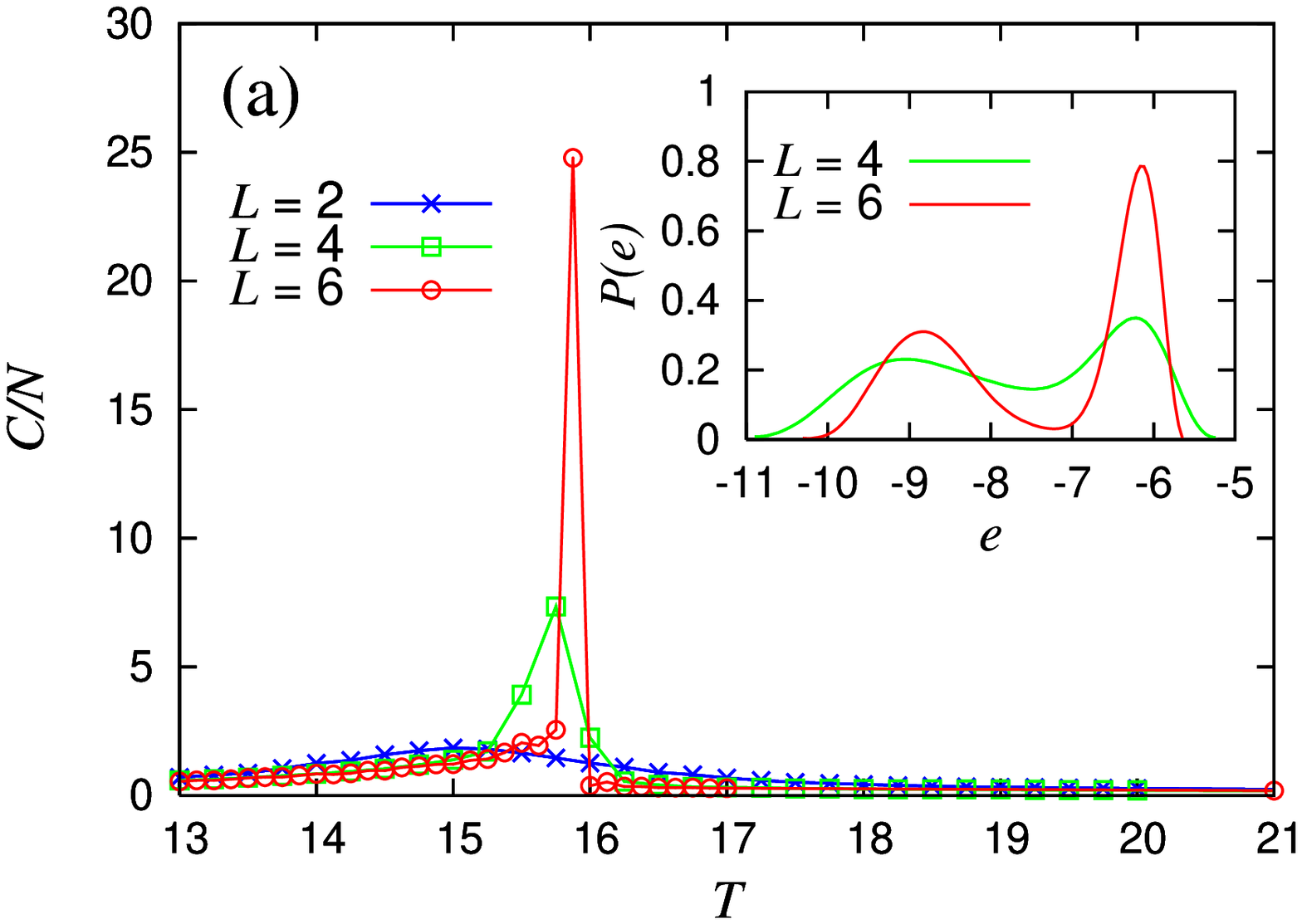}
\includegraphics[scale=0.4,angle=-90]{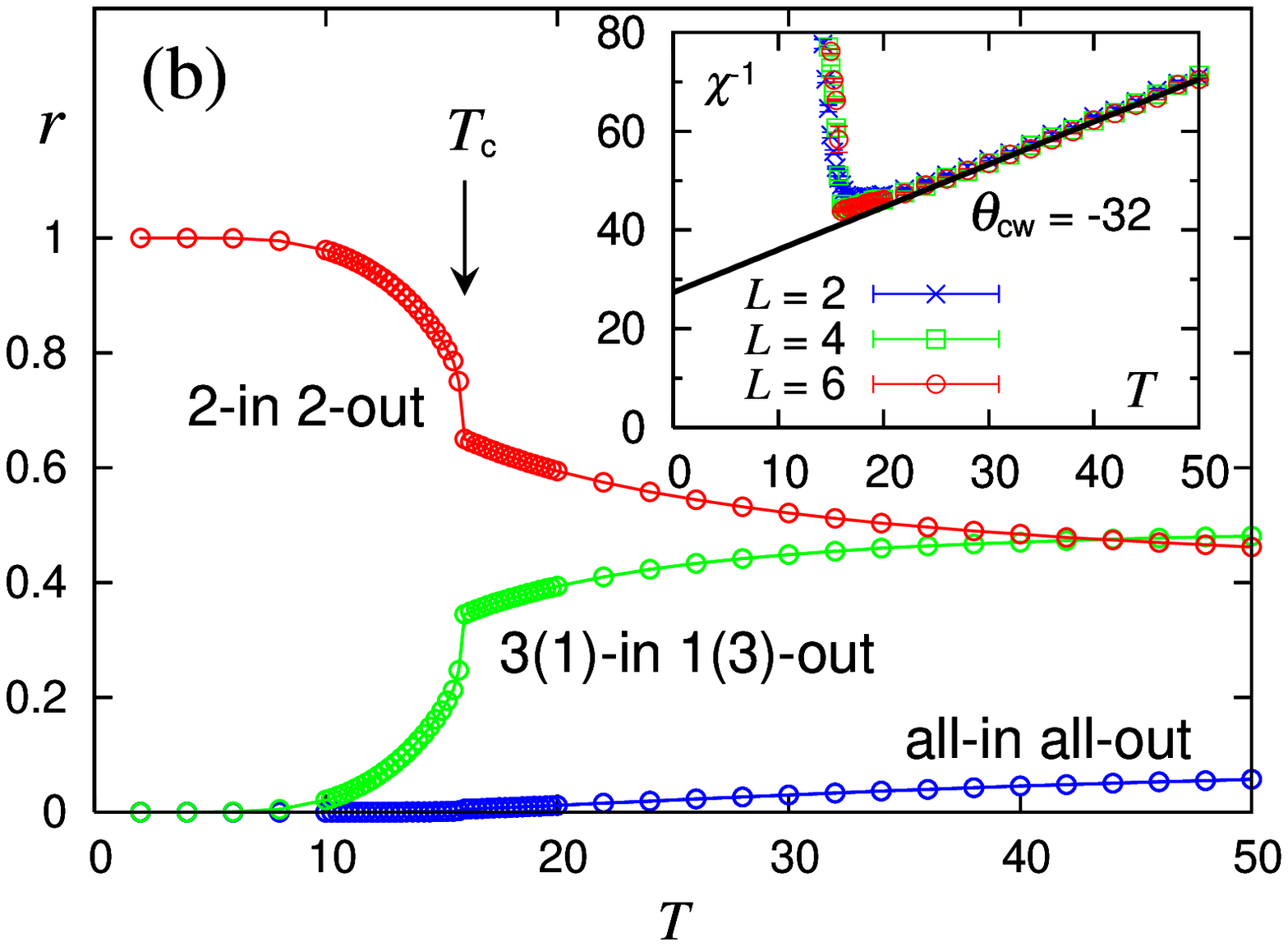}
\end{center}
\caption{
(Color online) The temperature dependence of the specific heat per spin (a), and of the temperature dependence of the ratio $r$ of the ``2-in 2-out'', ``3(1)-in 1(3)-out'' and ``all-in all-out'' local spin structures at each tetrahedron (b), for the parameter $k_F=2\pi /1.218$. In the inset of (a), the energy histogram is shown at a temperature close to the transition temperature, at $T=15.750$ ($L=4$) and at $T=15.875$ ($L=6$). The inset of (b) exhibits the inverse susceptibility per spin $\chi^{-1}$.
}
\end{figure}

 Naively, one may think that the antiferromagnetic $\Theta_{{\rm CW}}$ is associated with the ``all-in all-out'' local spin structure. In fact, this is not the case here. The inset of Fig.1(b) exhibits the temperature dependence of the ratios of the ``2-in 2-out'' structure, the ``3(1)-in 1(3)-out'' structure and the ``all-in all-out'' structure, which represent the probability that each type of local spin structure is realized at each tetrahedron. As can be seen from the figure, the local spin structure below $T_c$ is ``2-in 2-out'', not ``all-in all-out''. The appearance of the ferromagnetic  ``2-in 2-out'' spin structure under the antiferromagnetic $\Theta_{\rm CW}$ is consistent with the experimental observation for Pr$_2$Ir$_2$O$_7$ \cite{Nakatsuji}. In fact, as we shall see below, such an apparently contrasting behavior can be understood as a consequence of the long-range and oscillating nature of the RKKY interaction.

 To further clarify the nature of the ordered state, we calculate the spin  structure factor  $F(\vec{q})$, {\it i.e.\/}, the thermal average of the squared Fourier amplitude defined by
\begin{equation}
   F(\vec{q}) = \frac{1}{N} \bigl\langle \bigl| \sum_j \vec{S}_j~e^{i\vec{q}
   \cdot \vec{r}_j} \bigr|^2 \bigr\rangle, 
\end{equation}
where the wavevector is given by $\vec q=\frac{2\pi}{a}(h,k,l)$ with $-2\leq h,k,l\leq 2$. In Fig.2(a), we show the structure factor in the $(h,h,l)$ plane calculated at a low temperature $T=2$, which has been averaged over all equivalent directions of ($\pm h,\pm h,\pm l$), ($\pm l, \pm h, \pm h$) and ($\pm h, \pm l, \pm h$). As can be seen from the figure, the spin structure factor exhibits a sharp Bragg peak at $(0,0,1)$, or at the equivalent ones, indicating that the magnetic long-range order sets in below $T_c$, the associated spin order being characterized by the wavevector $(0,0,\frac{2\pi}{a})$. We have confirmed that the Bragg peaks appear at these spots irrespective of $L$, and sharpen with $L$. It is not a finite-size effect.

\begin{figure}[ht]
\begin{center}
\includegraphics[scale=0.45]{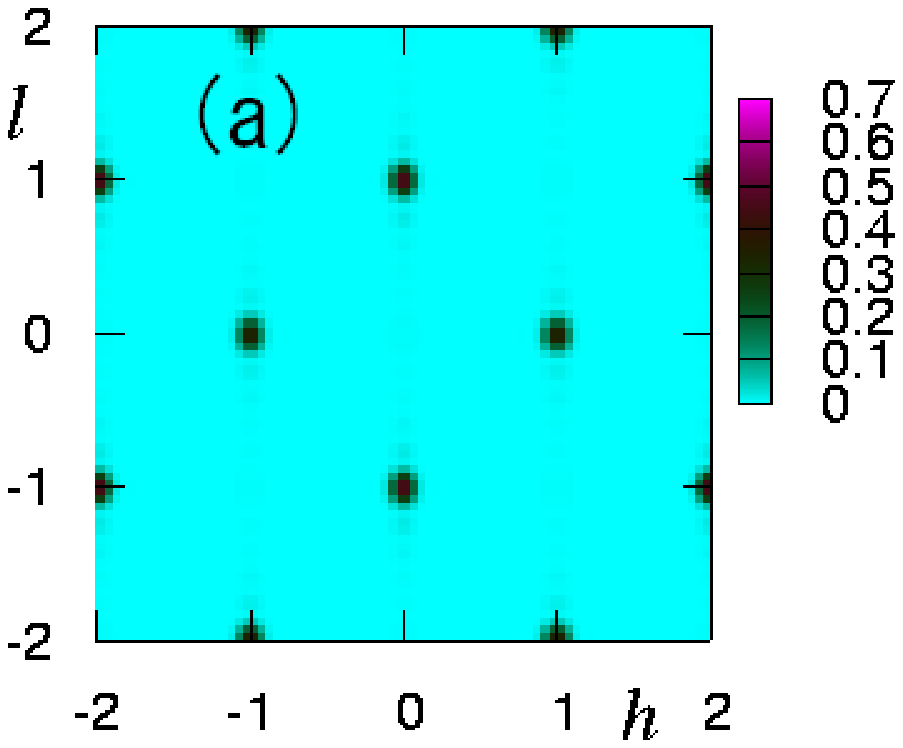}
\includegraphics[scale=0.45]{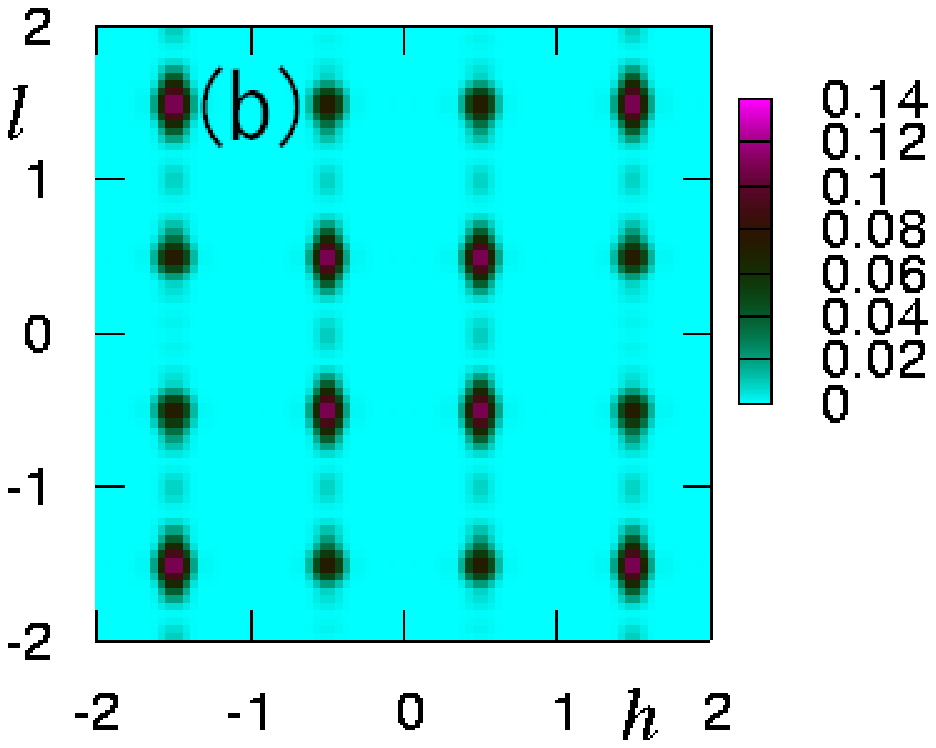}
\end{center}
\caption{
(Color online) The spin structure factor $F(\vec q)$ in the ($h,l$) plane at a temperature $T=2$ for $k_F=2\pi /1.218$ (a), and for $k_F=2\pi /1.13$ (b). The system size is $L=6$.
}
\end{figure}

 In fact, such an ordered state is the same as the one observed in  the dipolar spin-ice model in its equilibrium \cite{Melko}. In the case of the dipolar spin-ice model, the $(0,0,\frac{2\pi}{a})$ ordered state is observed only in a special type of simulation devised to promote equilibration (loop algorithm), not in the standard single spin-flip simulation \cite{Hertog}: In the latter case, the system gets out of equilibrium leading to the glassy, spin-ice behavior. In our present RKKY model, by contrast, the $(0,0,\frac{2\pi}{a})$ long-range order is achieved more readily, even without using the loop algorithm. 

 The $(0,0,\frac{2\pi}{a})$ ordered state is six-fold degenerate reflecting the cubic symmetry of the lattice. The order parameter possesses a structure equivalent to the one of the ferromagnetic cubic model. Since the latter model is known to exhibit a first-order transition in three dimensions \cite{cubic}, the observed first-order nature of the transition is naturally understood as a consequence of the occurrence of the $(0,0,\frac{2\pi}{a})$ ordered state below $T_c$. 

 In order to examine the stability of the ordering against the possible change of the parameter $k_F$, we also simulate the model with other values of $k_F$. For the wavevectors $k_F=2\pi /1.15$ and $2\pi/1.4$ close to the case studied above, we have observed essentially the same ordering behavior, {\it i.e.\/}, a first-order transition into the $(0,0,\frac{2\pi}{a})$ ordered state consisting of the ``2-in 2-out'' local spin structure, which is accompanied by the negative $\Theta_{\rm CW}$.

 If one further increases or decreases the $k_F$-value, somewhat different behaviors arise. With increasing the $k_F$-value, $|\Theta _{\rm CW}|$ decreases its magnitude, eventually changing its sign from negative to positive at around $k_F\simeq 2\pi /1.14$. At $k_F=2\pi /1.13$,  $\Theta_{\rm CW}$ becomes positive (ferromagnetic), while the model exhibits a first-order transition into the long-range ordered state with keeping the ``2-in 2-out'' local spin structure. The ordered state is characterized by the wavevector $(\frac{\pi}{a},\frac{\pi}{a},\frac{\pi}{a}$) which is distinct from the $(0,0,\frac{2\pi}{a})$ ordered state observed at $k_F=2\pi /1.218$. The computed spin structure factor at $k_F=2\pi /1.13$ is shown in Fig.2(b). 

 On the other hand, with decreasing the $k_F$-value from $k_F=2\pi /1.218$,  the sign of $\Theta _{\rm CW}$ is kept to be negative and the local spin structure remains to be ``2-in 2-out'', while at $k_F=2\pi /1.7$ the model exhibits a first-order transition into the more complex long-range ordered state characterized by the wavevectors with all three spin components different from each other, $h\neq k\neq l$. 

 In Fig.3, we summarize the ordering behavior of the model for a wider range of  $k_F$. The sign of the Curie-Weiss constant $\Theta_{\rm CW}$ determined from the computed $\chi^{-1}$ and the type of the local spin structure, {\it i.e.\/}, either ferromagnetic ``2-in 2-out'' or antiferromagnetic ``all-in all-out'', are shown. We emphasize that, at any $k_F$-value studied, we have never observed a spin-liquid-like behavior in which the spin long-range order is suppressed down to $T=0$ or temperatures much lower than $\Theta_{\rm CW}$. Further details will be reported elsewhere.

 The origin of the sign of the effective couplings can be understood as follows. First, the type of the local spin order, {\it i.e.\/}, either ``2-in 2-out'' or ``all-in all-out'', turns out to be primarily dictated by the sign of the {\it nearest-neighbor\/} part of the RKKY interaction. To see this, we include in Fig.3 the sign of the nearest-neighbor part of the RKKY interaction as a function of $k_F$, which turns out to reproduce the sign of the effective coupling associated with the local spin structure.

 By contrast, {\it distant-neighbor\/} interactions also contribute to its Curie-Weiss constant. This can be seen most easily from the lowest-order high-temperature expansion, which yields
\begin{equation}
k_B\Theta_{\rm CW} \simeq \sum_{j\in {\rm sub}(i)} J_{ij} + \frac{1}{9}\sum _{j\in {\rm sub}(i)} J_{ij},
\end{equation}
where the summation $j\in {\rm sub}(i)$ is taken over all sites $j$ belonging to the same fcc sublattice as the site $i$, while the summation $j\in {\rm sub}(i)$ taken over all sites $j$ belonging to other three fcc sublattices. The factor $1/9$ in the second term arises due to the mutual canting of the $<111>$ easy axes at different sublattices. The sign of eq.(4) is also shown in Fig.3.  As can be seen from the figure, it reproduces the sign of $\Theta _{\rm CW}$ determined from our simulation reasonably well. Thus, in case of the long-range interaction like the RKKY interaction,  the local spin structure at each tetrahedron is primarily dictated by its nearest-neighbor part, while the distant-neighbor interactions also contribute to the Curie-Weiss constant.

\begin{figure}[ht]
\begin{center}
\includegraphics[scale=0.3]{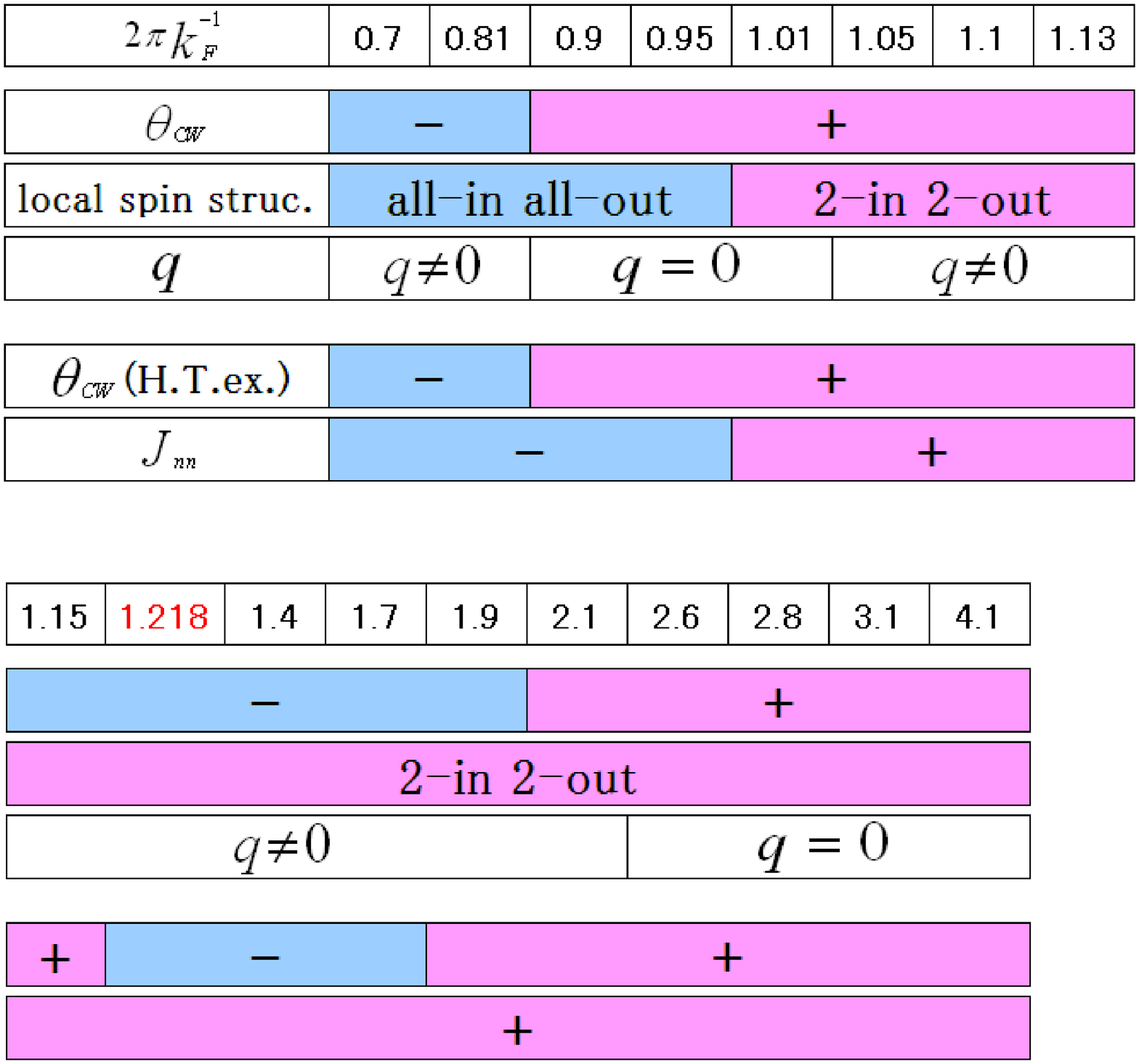}
\end{center}
\caption{
(Color online) The $k_F$-dependence of the sign of the Curie-Weiss constant $\Theta _{\rm CW}$ and the sign of the effective coupling associated with the type of the local spin structure, {\it i.e.\/}, either ferromagnetic ``2-in 2-out'' or antiferromagnetic  ``all-in all-out''. The information of the wavevector $q$ characterizing the ordered state, whether it is $q=0$ or $q\neq 0$, is also given. For comparison, the sign of the nearest-neighbor part of the RKKY interaction $J_{nn}$ and the sign of $\Theta _{\rm CW}$ calculated from the high-temperature expansion (H.T.ex.), eq.(4), are given. 
}
\end{figure}


\begin{figure}[ht]
\begin{center}
\includegraphics[scale=0.45]{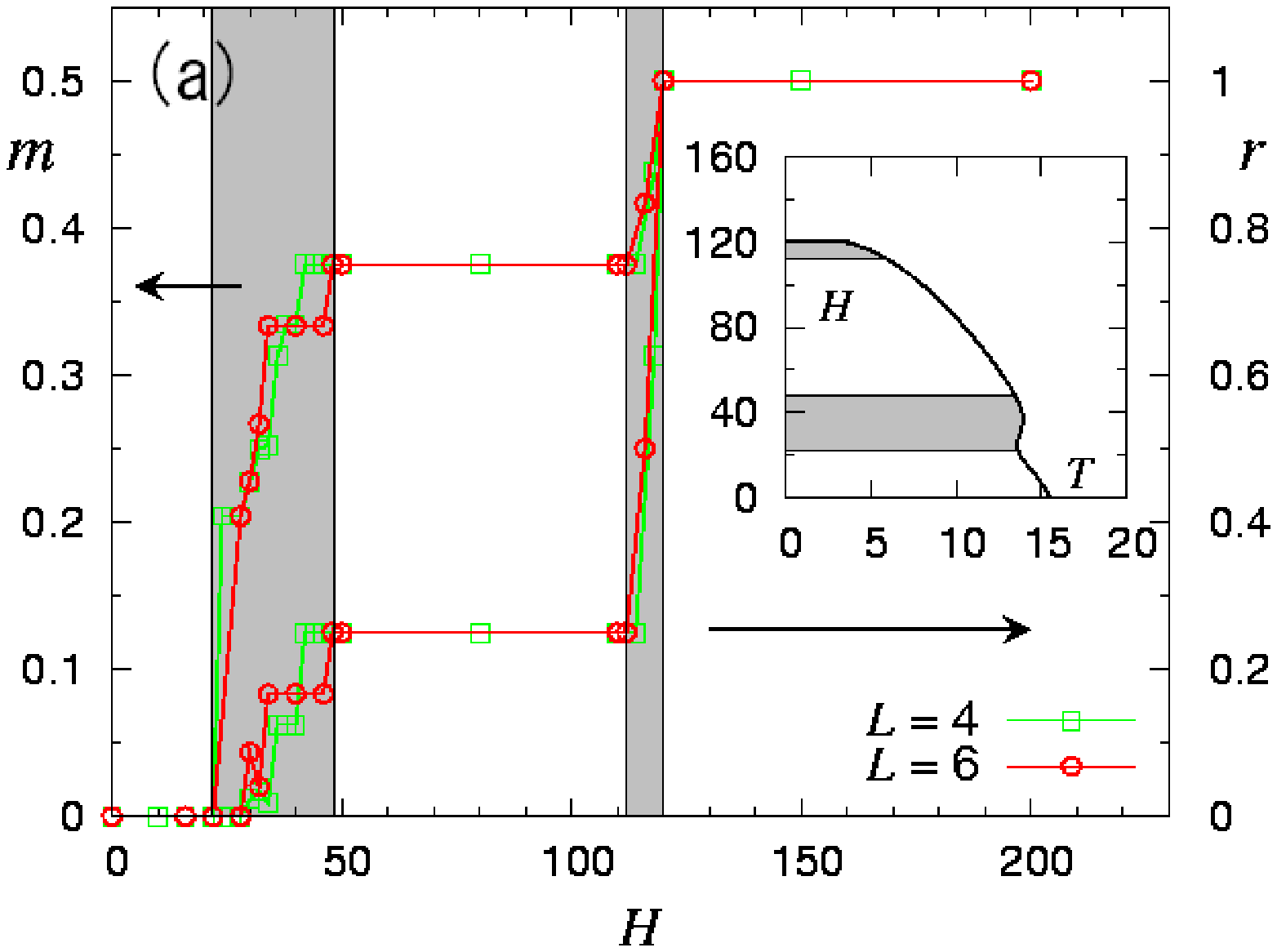}
\includegraphics[scale=0.26]{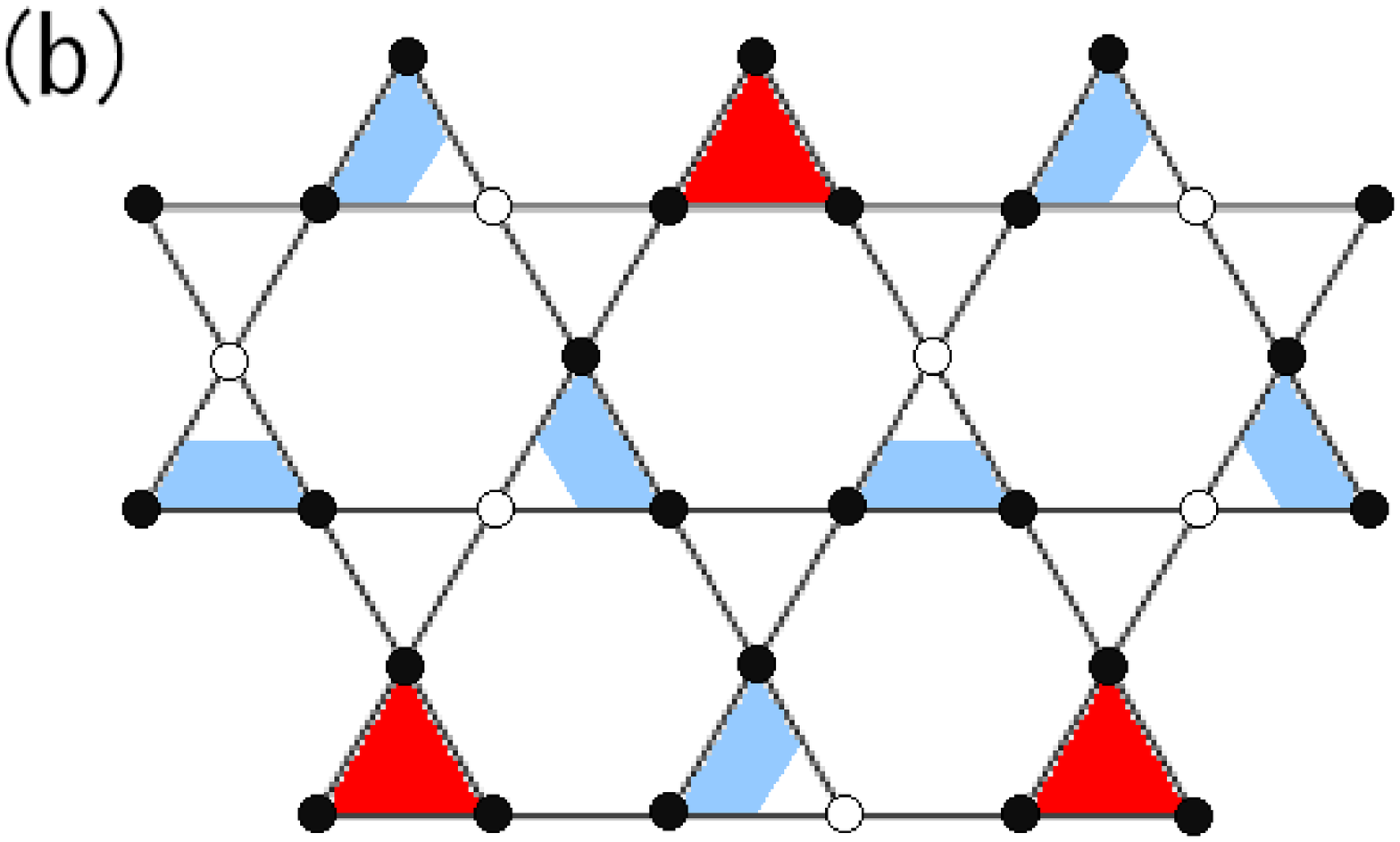}
\end{center}
\caption{
(Color online) (a) The magnetization per spin plotted versus the [111] field intensity at a low temperature $T=0.5$ for $k_F=2\pi /1.218$, together with the ratio $r$ of the ``3(1)-in 1(3)-out'' local spin structure. The lattice size is $L=4$ and 6. The inset is a phase diagram in the field-temperature plane. In the shaded region, the detailed phase structure cannot be identified due to strong finite-size effects. (b) Spin pattern on a (111) kagom\'e layer at the intermediate plateau state with $m=3/8$. The black (white) circle represents the ``in'' (``out'') spin state. The red or thick (blue or thin) triangle represents a face of the ``3-in 1-out'' (``2-in 2-out'') tetrahedron.
}
\end{figure}

  In the case of dipolar spin ice, the in-field properties turned out to be quite rich and interesting \cite{Ruff}. In view of this, we also investigate the ordering behavior of our RKKY model under magnetic fields applied along the [111] direction for the ``experimental'' case of $k_F=2\pi/1.218$.  The calculated magnetization per spin $m$ is shown in Fig.4(a). In addition to the low-field plateau at $m=0$ corresponding to the $(0,0,\frac{2\pi}{a})$ ordered state and the high-field one at $m=1/2$, there appears for intermediate fields  another plateau at $m=3/8$. Note that the pyrochlore lattice can be viewed as an alternate stack of a sparse triangular layer and a kagom\'e layer. Under sufficiently strong [111] magnetic field, all tetrahedra take ``3(1)-in 1(3)-out'' configurations where all spins on a triangular layer are aligned parallel with the [111] field, with the saturation magnetization $m=m_s=1/2$. In the intermediate plateau state, a quarter of tetrahedra takes ``3(1)-in 1(3)-out`` configuration, while other 3/4 take ``2-in 2-out'' configuration: See Fig.4(a). These ``3-in 1-out'' tetrahedra are arranged periodically on the pyrochlore lattice, {\it i.e.\/},  these tetrahedra form a $2\times 2$ triangular superlattice on a (111) kagom\'e layer as depicted in Fig.4(b). In the inset of Fig.4(a), we sketch a phase diagram of the model for $k_F=2\pi /1.218$ in the field-temperature plane. In the shaded region of the phase diagram, due to the possible the incommensurability effect, complicated behaviors accompanied by various narrow plateaus are observed, which, however, are strongly size-dependent in the range of small sizes studied here.

 Finally, on the basis of our present finding for the RKKY pyrochlore Ising model, we wish to discuss recent experiments on metallic pyrochlores. Our present result is consistent with the experimental result on Pr$_2$Ir$_2$O$_7$ in that the local spin structure is ``2-in 2-out'' whereas the Curie-Weiss temperature is antiferromagnetic \cite{Nakatsuji}. By contrast, a first-order transition into the $(0,0,\frac{2\pi}{a})$ ordered state as observed in our simulation is not observed experimentally. If one deduces the transition temperature by fitting the calculated $\Theta_{\rm CW}$ to the experimental $\Theta_{\rm CW}$, one gets an estimate of $T_c\simeq 10$K. Experimentally, below 10K, Pr$_2$Ir$_2$O$_7$ gets into the Kondo regime. This coincidence strongly suggests that the hybridization effect between the itinerant Ir conduction electrons and the localized Pr moments, not just the effect of magnetic frustration, is essential in realizing the spin-liquid-like behavior observed in Pr$_2$Ir$_2$O$_7$ at low temperatures. Meanwhile, small difference in samples may possibly stabilize the type of long-range ordered state revealed here. Indeed, a series of pyrochlore magnets, {\it e.g.\/}, pyrochlore iridates $R_2$Ir$_2$O$_7$ including both metallic and insulating, exhibit a variety of ordering behaviors \cite{Yanagishima,Matsuhira}, and our present result on the RKKY pyrochlore Ising model might serve as a useful reference in interpreting the experimental data.

 The authors are thankful to Dr. S. Nakatsuji, Dr. Y. Machida, Dr. S. Onoda, Dr. M. Gingras and Dr. C. Broholm for useful discussion. This study was supported by Grant-in-Aid for Scientific Research on Priority Areas ``Novel State of Matter Induced by Frustration'' (19052006).


\begin{thebibliography}{30}

\bibitem{Harris} 
M. J. Harris, S. T. Bramwell, D. F.. McMorrow, T. Zeiske, and
K. W. Godfrey, Phys. Rev. Lett. {\bf 79}, 2554 (1997).

\bibitem{Ramirez} 
A.P. Ramirez, A. Hayashi, R.J. Cava, R. Siddharthan, and B.S. Shastry,   
Nature {\bf 399}, 333 (1999).

\bibitem{Bramwell}
 S.T. Bramwell and M.J.P. Gingras, Science {\bf 29}, 1495 (2001).

\bibitem{review}
For review, see R.G. Melko and M.J.P. Gingras, J. Phys. Condens. Matter {\bf 16}, R1277 (2004).

\bibitem{Siddharthan}
R. Siddharthan, B.S. Shastry, A.P. Remirez, A. Hayashi, R.J. Cava and S. Rosenkranz, Phys. Rev. Lett. {\bf 83}, 1854 (1999).

\bibitem{Hertog}
 B.C. den Hertog and M.J.P. Gingras, Phys. Rev. Lett. {\bf 84}, 3430 (2000).

\bibitem{Nakatsuji}
S. Nakatsuji, Y. Machida, Y. Maeno, T. Tayama, T. Sakakibara, J. van
Duijn, L. Balicas, J. N. Milican, R. T. Macaluso, and Julia Y. Chan,  
Phys. Rev. Lett. 96 087204 (2006); Y. Machida, S. Nakatsuji, Y. Maeno, T. Tayama, T. Sakakibara, and S. Onoda, Phys. Rev. Lett. {\bf 98}, 057203 (2007).

\bibitem{Hansen}
J.P. Hansen,  Phys. Rev. A{\bf 8}, 3096 (1973).

\bibitem{Melko}
R.G. Melko, Byron C. den Hertog, and M.J.P. Gingras, Phys. Rev. Lett. {\bf 87}, 067203 (2001).

\bibitem{cubic}
E. Nienhuis, E.K. Riedel and M. Shick, Phys. Rev. B{\bf 27}, 5625 (1983). 

\bibitem{Ruff}
 See, {\it e.g.\/}, J.P.C. Ruff, R.G. Melko and M.J.P. Gingras, Phys. Rev. Lett. {\bf 95}, 097202 (2005). 

\bibitem{Yanagishima}
D. Yanagishima and Y. Maeno, J. Phys. Soc. Jpan. {\bf 70}, 2880 (2001).

\bibitem{Matsuhira}
K. Matsuhira, M. Wakeshima, R. Nakanishi, T. Yamada, A. Nakamura, W. Kawano, S. Takagi and Y. Hinatsu, J. Phys. Soc. Jpan. {\bf 76}, 043706 (2007); N. Taira, M. Wakeshima and Y. Hinatsu, J. Phys. Condens. Matter {\bf 13}, 5527 (2001).





\end{thebibliography}
\end{document}